\documentclass[a4paper,american,11pt]{scrartcl}

\usepackage[top=1in,bottom=1in,left=1in,right=1in]{geometry}

\usepackage{lmodern}
\usepackage[utf8]{inputenc}
\usepackage[T1]{fontenc}
\usepackage{color}
\usepackage{todonotes}
\usepackage{babel}
\usepackage{amsmath}
\usepackage{bbm}
\usepackage{amssymb}
\usepackage{setspace}
\usepackage[authoryear]{natbib}
\usepackage[font=small]{quoting}
\usepackage{latexsym}
\usepackage[bookmarks=false,pdfborder={0 0 0},colorlinks=false]{hyperref}
\usepackage{enumerate}
\usepackage[normalem]{ulem}



\begin{document}

\title{A persistent particle ontology for QFT in terms of the Dirac sea}

\author{Dirk-André Deckert%
\thanks{Ludwig-Maximilians-Universität München, Mathematisches Institut,
Theresienstrasse 39, 80333 München, Germany. E-mail:
\protect\href{mailto:deckert@math.lmu.de}{deckert@math.lmu.de}}%
, Michael Esfeld%
\thanks{Université de Lausanne, Section de philosophie,
1015 Lausanne, Switzerland. E-mail: \protect\href{mailto:Michael-Andreas.Esfeld@unil.ch}{Michael-Andreas.Esfeld@unil.ch}%
}, Andrea Oldofredi%
\thanks{Université de Lausanne, Section de philosophie,
1015 Lausanne, Switzerland. E-mail: \protect\href{mailto:Andrea.Oldofredi@unil.ch}{Andrea.Oldofredi@unil.ch}}
}

\maketitle
\begin{abstract}
      \begin{center} Forthcoming in the \emph{British Journal for the Philosophy of Science}    \end{center}
    \medskip

We show that the Bohmian approach in terms of persisting particles that
    move on continuous trajectories following a deterministic law can be
    literally applied to QFT. By means of the Dirac sea model -- 
    exemplified in the electron sector of the standard model neglecting
    radiation -- we explain how starting from persisting particles, one is led
    to standard QFT employing creation and annihilation operators when tracking
    the dynamics with respect to a reference state, the so-called vacuum. Since
    on the level of wave functions, both formalisms are mathematically
    equivalent, this proposal provides for an ontology of QFT that includes a
    dynamics of individual processes, solves the measurement problem and
    explains the appearance of creation and annihilation events.

    \medskip{}

\end{abstract}

\tableofcontents{}


\section{Bohmian mechanics from QM to QFT}
\label{sec:introduction}

Bohmian mechanics (BM) is by now a well-respected proposal for the ontology of
non-relativistic quantum mechanis (QM). Any such proposal has to provide a
solution to the measurement problem by setting out what happens in nature on the
level of individual quantum systems instead of merely making statistical
predictions for measurement outcomes. BM meets this requirement by supplementing
the wave equation with a guiding equation that yields trajectories for
individual particles in three-dimensional space (what is known as the primitive
ontology) and that explains the states of macroscopic systems as
well as their stability in terms of these trajectories (see \cite{Durr:2013aa}). 

When it comes to QFT, the requirement for an ontology is the same as in QM: the
measurement problem plagues QFT in the same way as QM (see
\cite{Barrett:2014aa}); any proposal for its solution has to explain what
happens in nature on the level of individual processes instead of merely making
statistical predictions. On the one hand, a particle ontology is generally
regarded as excluded for QFT: if there is not a definite number of particles that persist, particles cannot be what is fundamental (see e.g. \cite{Clifton:2002aa}; \citet[][chapter 8]{Kuhlmann:2010aa};
\citet[][chapters 9-11]{Ruetsche:2011aa}). On the other hand, it is by no means
clear that the ontology of QFT is one of fields, since QFT does not admit fields
that have definite values at the points of four-dimensional space-time (see \cite{Baker:2009aa}).

Against this background, our claim is that the Bohmian approach works for QFT in the same way as for QM: as it is a \emph{non sequitur}
to take particle trajectories to be ruled out in QM due to the Heisenberg
uncertainty relations, so it is a \emph{non sequitur} to take permanent
particles moving on definite trajectories according to a deterministic law to be
ruled out in QFT due to the statistics of particle creation and annihilation
phenomena. In both cases, such an underlying particle ontology is in the
position to explain the statistics of measurement outcomes. Our claim therefore
is that if the Bohmian approach is a serious contender for the ontology of QM,
then it is also a serious contender for the ontology of QFT. We only seek to
establish this conditional claim. We are not concerned here with defending the Bohmian one against rival approaches.

Bohm himself in his
\cite{Bohm:1952ab}, appendix A, already
discussed a possible extension of his theory to the electromagnetic field.
Indeed there is a field version of Bohmian QFT (see \cite{Struyve:2006aa, Struyve:2007aa} as well as  \cite{Valentini:1992aa}, chapter 4). As regards the particle ontology, the first Bohmian QFT can
be traced back to \cite{Bell:1986aa} (see \cite{durr_bell-type_2005} for a
continuum version of Bell's proposal on a lattice; see
\cite{Struyve:2010aa} for an overview of all the Bohmian proposals). Bell's
proposal grants the entities that figure in the particle-pair creation and
annihilation processes of QFT the status of real, fundamental particles,
which are, accordingly, created and annihilated at random times and
positions. Between these random events, the particles evolve according to a
Bohmian law of motion.

We do not adopt this proposal. In the first place,
since these entities depend on the choice of an initial reference vacuum
state that is not unique (see \cite{fierz_particle_1979}), instead of being
objects that simply exist, we consider it to be inappropriate to grant them
the status of fundamental objects in the ontology.  More generally
speaking, the Bohmian approach is motivated by finding an underlying
ontology that explains the statistics of measurement outcomes. We think
that it is worth while to pursue this approach only if one is prepared to
go all the way down to an ontology of fundamental objects that are simply
there -- that is, that do not come into existence out of nothing and that
do not disappear into nothing -- and that evolve according to a simple,
deterministic law (that is, an evolution not interrupted by random jumps).

That is why we take up an idea going back to \cite{dirac_theorie_1934}: there
is no particle creation or annihilation. There are only conditions under which
particle motion becomes observable or fails to be so. These conditions are not
unique; they may even depend on the state of motion of the observer (as in the
Unruh effect). We build an ontology of permanent particles on this idea. In
this task we face a shortcoming of the modern formulation of QFT, which is
given only in terms of a scattering theory. The mathematical difficulties
involved in the computation of scattering matrix elements are already so severe
that in the renormalization group formulation of QFT, the question about a
fundamental equation of motion is hardly thematized. In ontology, however, we
have to make explicit what the laws are according to which individual processes
occur; for what there is in nature are individual events and processes.
Techniques developed to calculate measurement outcome statistics do not reveal
how these processes evolve from an initial to a final state.

We therefore propose to return to the early attempts of the 1930s to define
equations of motion for QFT. In doing so, we will not obtain better
predictions of measurement outcomes than standard QFT. The situation is the same
as in BM: the guiding equation does not yield better predictions; accordingly,
it does not have to be solved to calculate statistical predictions of
measurement outcomes. It figures in the explanation of these
statistics by answering the question of what is really going on ontologically.
Hence, the purpose in returning to \cite{dirac_theorie_1934} is not to pursue
outdated physics, but to make progress with respect to the ontology of QFT.
In
modern formulations of QFT, a mathematical description in terms of the Dirac
sea, although being canonically equivalent on the level of wave functions, has
been abandoned in favour of a more economic description involving particles and
anti-particles as well as their creation and annihilation. It is not our
intention to reintroduce the Dirac sea as computational device. On the
contrary, we will argue that after it served its ontological purpose, it
suggests itself to develop computational methods that let go the bulk of the
Dirac sea, replacing it by a so-called vacuum state, and only track its
excitations. What one gains in formulating the fundamental
equations of motion in terms of the Dirac sea is an ontological
explanation of the physical processes behind particle and anti-particle
creation and annihilation.

In the following, we examine the so-called \emph{Dirac sea model} or \emph{hole
theory}, which was the first one to predict the phenomenon of electron-positron
pair-creation. Although it is conceived only for the electron sector of the standard model (SM)
of QFT, it is applicable to all fermionic matter. As \cite{Bell:1986aa} argued, the commitment to fermions is sufficient to
account for the empirical data: they all consist in a spatial arrangement of
fermions. Hence, if the Dirac sea model can cover all fermionic matter, this is
all that can be reasonably demanded for it to serve as a proposal for the
ontology of QFT. Furthermore, this model can be given a rigorous mathematical
treatment in certain regimes. It is therefore suitable for probing the compatibility of QFT with an ontology of persistent
particles. \cite{Colin:2004aa} and \cite{Colin:2007aa} have carried out
an investigation in that sense in the physics literature (see also the brief
remarks in \citet[][pp. 373-374]{Bohm:1987aa} and \citet[][chapter
12.3]{Bohm:1993aa}). In the philosophical literature, the Dirac sea model is
hitherto largely ignored, apart from a brief assessment in \citet[][pp. 78-79,
86-88]{Saunders:1999aa}.
 
In the next section, we introduce the Dirac sea model (section~\ref{sec:model}).
Then, we define the state of equilibrium for the particles -- that is, a state
in which nothing can be observed, by physicists usually referred to as the
``vacuum state'' (section~\ref{sec:dirac-sea}) -- and small excitations from equilibrium, which can be
observed and which appear as if there were a creation and annihilation of
particles. This then is the basis on which we show
how the formalism that effectively describes these excitations leads naturally
to the standard QFT formalism, which employs creation and annihilation
operators. We discuss how, due to the ontology of persistent particles, it is
possible to give a clear meaning to these operators as well as to the vacuum
state (sections~\ref{sec:excitations} and~\ref{sec:formalism}). We conclude with
showing how this ontology and dynamics explains the measurement outcomes,
arguing that it does for QFT what BM does for QM.

\section{The Dirac sea model}
\label{sec:model}

The Dirac sea model can be inferred from the SM by imposing the following
restrictions:
\begin{enumerate}
    \item Restriction to the electron sector of the SM;
    \item Restriction to direct electrodynamic interaction and neglect of
        radiation;  
    \item Modelling interaction with all other fermion sectors of the
        SM only effectively by a time-dependent ``external''
        interaction.
\end{enumerate}
The resulting model is simple enough to enable an unobscured
discussion, but has sufficient structure to describe the phenomenon of
electron-positron pair-creation. It therefore serves our purpose well. In
posing these restrictions, we assume that the particles can be divided into two
groups: the first group -- below labelled by $1,\ldots, N$ -- are
electrons, while the rest are other fermions. Nonetheless, we emphasize again
that the restriction to electrons is arbitrary, since the Dirac sea model and
hence our argument is also applicable to the other fermion sectors and the other
types of interaction in the SM.

Electrons in direct electrodynamic interaction exclusively repel each other. If
this were the only interaction, the spatial extension of this cloud of $N$
electrons would inevitably grow larger and larger without any bound. This is an
artifact of restrictions 1-3, as we model only the electron sector directly 
(though we allow an indirect, effective interaction to other fermion sectors). To avoid such
unphysical behaviour we stipulate further that, due to the motion of all the
other particles and their interaction by means of an ``external interaction''
(which then includes also attractive interactions, even gravitation, although
the latter is not described by the SM), the spatial extension will always be
bounded (which might also be suggested by general relativity and supported by
cosmological evidence). Furthermore, we assume that this ``external'' interaction
is reasonably well behaved in the following sense: neither will it dampen the
motion of the electrons to such an extent that all electron motion comes to a
rest, nor will it drive the electron velocities arbitrarily close to the speed
of light. Thus, there is no infinite energy transfer. The idea behind this
assumption is that motion should be somewhat conserved among all fermion
sectors. Neither does motion arise from nothing nor does it cease to exist, it
only varies over the individual particles.

Posing these two additional restrictions allows us to avoid the
discussion of mathematical problems -- notably the infamous ultraviolet / infrared
divergences --, which are the obstacle in finding a well-defined equation of
motion in QFT. Put mathematically, we assume that the initial data and the
``external'' interaction in our theory are such that:
\begin{enumerate}[1.]\setcounter{enumi}{3}
    \item The spatial extension of the universe is restricted to a finite volume
        $\Gamma$ in 3d-space $\mathbb R^3$.
    \item The electron momenta are restricted to be lower than some finite
        ultraviolet cut-off $\Lambda$. 
\end{enumerate}
In a yet to be found well-defined SM of QFT, the above assumptions
must not be imposed \emph{a priori}, but have to come out \emph{a posteriori}.
In particular, the gained mathematical well-definedness comes at
the cost of a violation of Lorentz invariance and the introduction of seemingly
arbitrary parameters. Accordingly, the main objective of modern renormalization theory is the
removal of these unwanted cut-offs. As yet, however, a removal is only achieved
order by order in the informal expansions of scattering matrix elements. When it
comes to an equation of motion, the introduction of cut-offs is still
standard in QFT. Nevertheless, it is the hope amongst physicists that
provided the cut-offs are large enough, the effects on the dynamics are only
small in certain regimes of interest. In fact, given the cut-offs, the
formulation presented here is equivalent to the textbook formulation on the
level of wave functions. By introducing another cut-off in the
respective Bose fields, it is furthermore possible to include the full
electroweak interaction including radiation effects and remove assumptions
1-3 altogether (see \cite{Colin:2007aa}). In general, the Dirac sea model applies to any fermion sector and
potentially any interaction between those fermions (except maybe
gravitation). In short, the above assumptions
are no objectionable limitations of our endeavour. They rather underline the
general point that the rationale of the Bohmian approach is not to obtain better
physics, but an ontology for the existing physics.

The Dirac sea model was originally proposed by
\cite{dirac_theorie_1934,Dirac:1947} in Hartree-Fock approximation in terms of
density matrices. Without this approximation it was later included in QFT by
means of a second quantization procedure of the one-particle Dirac equation
yielding the so-called Dirac field operators on Fock space. We discuss the Dirac sea model in terms of a straightforward quantum mechanical
$N$-particle wave equation. For finite $\Gamma$ and $\Lambda$, this formulation is
equivalent to the textbook version (see section~\ref{sec:excitations}). Additionally, we conceive the Dirac sea
model in terms of BM.

The Bohmian velocity field $v_t$ guiding the motion
of $N$ particles can be given in the form
\begin{gather}
    \label{eq:vel-field} 
    v_t(X) 
    =
   c\left( 
    j^{(k)}_t(X)/\rho_t(X)
    \right)_{k=1,\ldots,N},
    \\
    \rho_t(X) = \Psi_t(X)^* \Psi_t(X),
    \quad
    j^{(k)}_t(X) = \Psi_t(X)^* 1^{\otimes(k-1)}\otimes \alpha \otimes 1^{\otimes
    (N-k)}\Psi_t(X)  ,
\end{gather} 
where $X=(x_1,\ldots,x_N)\in\Gamma^N$, and $\rho_t$ and $j_{t}^{(k)}$ for
$k=1,\ldots,N$ are the probability density and currents generated by a wave
function $\Psi_t$ and $c$ stands for the speed of light. Here, $\Psi_t$ denotes
an anti-symmetric, square-integrable, $N$-particle spinor-valued function on
configuration space $\Gamma\subset\mathbb R^{3N}$, in short $\Psi_t\in\mathcal H^{\wedge N}$
for  $\mathcal H=L^2(\Gamma,\mathbb C^{4})$, that solves the wave equation
\begin{gather} 
    i\hbar\partial_t
    \Psi_t(x_1,\ldots,x_N) 
    =
    H^N
    \Psi_t(x_1,\ldots,x_N),
    \nonumber
    \\
    H^N
    =
    \sum_{k=1}^N \left(
    H_k^0(x_k)+ V_k(t,x_k) + H^I_k(x_k) \right).
    \label{eq:dirac-eq} 
\end{gather} 
The Hamiltonian $H^N$ is made of the free
Dirac Hamiltonian $H_0^k(x)$; the external influences $V_k(x)$ are given
by
\begin{align}
    \begin{split}
        H^0_k(x)
        & =
        1^{\otimes (k-1)}\otimes H^0(x)\otimes 1^{\otimes (N-k)},
        \\
        V_k(t,x)
        & =
        1^{\otimes (k-1)}\otimes V(t,x)\otimes 1^{\otimes (N-k)},
    \end{split}
    \begin{split}
        &H^0(x)
        = 
        -ic\alpha\cdot \nabla_x +\beta m c^2,
        \\
        \qquad
        &\text{for some external potential $V(t,x)$,}
    \end{split}
    \label{eq:H0}
\end{align} 
 where $m$ stands for the electron mass and
$\otimes$ denotes the tensor product (note that $\Psi$ is spinor-valued).
Furthermore, $\beta$ and the components of the vector
$\alpha=(\alpha_1,\alpha_2,\alpha_3)$ are $\mathbb C^{4\times 4}$ matrices that
fulfill the anti-commutator relations
$\{\alpha_j,\alpha_k\}=0=\{\beta,\alpha_j\}$ and $\beta^2=1=\alpha_j^2$ for
$j\neq k$, $j,k\in\{1,2,3\}$. Moreover, $H_I$ denotes the Hamiltonian modelling
the interaction between the particles. Since we neglect radiation, the
interaction is mediated directly by 
\begin{align}
    \label{eq:HI}
    H^I_k(x)
    &=\frac12 \sum_{j\neq k} U(x-x_j),
\end{align}
where the electric interaction between the $N$ electrons can be taken to be the
Coulomb potential $U(x)=\frac{e^2}{4\pi \epsilon_0}|x|^{-1}$. Here, $\epsilon_0$
is the electric constant and $e<0$ the charge of an electron. Note that $H^I_k$
depends also on $x_j$, which is suppressed in our notation. Finally, the
composite, effective interaction of all particles on electron $k$ is given
through a time-dependent potential $V_k(t,x)$, stemming for example from the
Coulomb field of a present ion, etc.   

In sum, the velocity field \eqref{eq:vel-field}
and the wave equation \eqref{eq:dirac-eq} define what we call the Dirac sea model. Given an initial configuration $Q_0$ and wave function
$\Psi_0$, the corresponding solutions of these equations yield a
unique trajectory of configurations of the $N$ electrons 
parametrized by time. In principle, there is nothing more to say
about the ontology: there are $N$ persistent electrons that evolve according to a deterministic law of motion. However, we face the following situation:
\begin{enumerate}[(I)]
    \item Generically, we do not have complete knowledge about the initial data
        $(Q_0, 
        \Psi_0)$.
    \item Even if we did, if $N$ is large (which is the interesting case,
        as we assume large cut-offs), it is in general neither
        analytically nor numerically feasible to compute solutions $\Psi_t$ of the
        wave equation \eqref{eq:dirac-eq}.
\end{enumerate}

Problem (I) is generic to any theory that applies to the universe as a whole.
Predictions about subsystems have to be inferred from a detailed statistical
analysis of what is to be expected in most situations with respect to a
meaningful measure. In BM, this is made possible by the special form of the
velocity law \eqref{eq:vel-field}, which ensures that if the initial data
$Q_0$ is distributed at random according to $|\Psi_0(X)|^2 d^{3N}X$,
configuration $Q_t$ is distributed according to $|\Psi_t(X)|^2 d^{3N}X$. This
feature is known as equivariance. The statistical analysis of subsystems has been
carried out in \citet[][chapter 2]{Durr:2013aa} for BM.
This analysis applies to our setting as well. The bottom line is a proof that
Born's rule holds true: if the effective wave function of the subsystem is
given by $\phi_t(x_1,\ldots,x_n)$ for $n\leq N$, then the distribution of the
subconfiguration of particles $q(t)=(q_1(t),\ldots,q_n(t))$ is given by
$|\phi_t(x)|^2 d^nx$.

Problem (II) stems from the fact that the pair potential $U$ strongly
entangles all
tensor components of the wave function $\Psi_t$ during the time evolution. Even
a perfect initial anti-symmetric product state will therefore
immediately lose its product structure due to \eqref{eq:dirac-eq}. The
complexity of this entanglement increases exponentially with $N$. Consequently,
one must hope to find special non-trivial situations in which the very
complicated $N$-particle dynamics becomes simple enough so that it can be
approximated by studying the motion of a few particles only.

The original motivation for Dirac's hole theory was not the complexity
of its solutions, but stemmed from the attempt to make sense of Dirac's equation
describing only one particle (see \cite{Saunders:1991aa} for the context of
Dirac's theory). One quickly realized that the solutions showed strange
behaviour -- for instance the \emph{Zitterbewegung} (\cite{schroedinger:1930})
and \emph{Klein's paradox} (\cite{Klein:1929}) --, which made Dirac's equation
hard to interpret physically. The mathematical reason for this strange behaviour
is the presence of a negative energy spectrum of the free Dirac Hamiltonian
$H^0(x)$ as given in \eqref{eq:H0}. Therefore, the starting point of Dirac's
hole theory was to somehow suppress this negative spectrum. As the electrons
are fermions, this can be accomplished by filling all the negative energy
states with a sea of particles -- and, granted the cut-offs 4-5, thus ending
up with an $N$-particle wave equation as above. The Pauli exclusion
principle then prevents the wave function of any additional electron from
growing negative energy components, because all negative energy states are
already occupied. The consequence is of course that one ends up with the same
complexity problem (II) as well, albeit for a different reason. 

Our starting point is different. We seek a theory about the universe as a whole,
based on an ontology of permanent particles. We bring in Dirac's hole theory as
a suitable means to implement that ontology. This model together with its
ontology is spelled out concisely in this and the preceding section. However, even though we only treat a
simplified model of that theory and only describe electrons while modeling the
rest of the particle interactions effectively, it is natural to start with not
only one but all the electrons of the universe. The model can thus only be
considered as a serious contender, if it is also possible
to analyze it mathematically despite of the complexity problem (II), which is
why we anyhow have to cope with a large number of electrons $N$. This is the
content of the following sections: our aim is to show how on the basis of this
model we can explain the statistical predictions of standard QFT.

Before doing so, let us compare our approach with the discussion of the Dirac
sea model in a Bohmian framework by \cite{Colin:2007aa}. They also assume a
finite fermion number; they are committed to positions for fermions for both
positive and negative energy particles, whereas there are no bosons in the
ontology. Anti-particles are defined as holes in the sea of the negative energy
particles. Whereas \cite{Colin:2007aa} define the fermion number as the total
number of particles minus the number of negative energy particles plus an
infinite constant, we are committed to \emph{N} particles that de facto
coincide with the fermions. Nonetheless, even though in our model there is no
ontologically significant difference between positive and negative energy
particles, one may say with the usual jargon that the fermion number can be
defined as the number of positive energy particles plus the number of negative
energy particles, which remains constant. \cite{Colin:2007aa} propose an
equation of motion for fermions that defines the vector velocity field for
configurations of particles, which is dependent on the wave function of the
system according to the usual Bohmian recipe; the expectation value of the
fermion number density is related to the position density: intuitively, the
number of fermions in a given region of space corresponds to the fermion
positions in that region. This model is regularized with the introduction of
ultraviolet momentum cut-offs and finite space; these constraints ensure that
it is a mathematically well defined theory.  Equivariance guarantees that the
empirical predictions of the SM with cut-offs are reproduced, exactly as
in our case. 

We emphasize again that the SM only permits to remove the introduced cut-offs
when computing perturbative corrections of scattering amplitudes with respect
to non-interacting QFT, while so far the respective dynamics of the SM also
becomes ill-defined with any attempt to remove the cut-offs. Since both the
theories of \cite{Colin:2007aa} and the one developed in this paper are
concerned with the dynamics and not with scattering theory, they can only be
compared to the SM with cut-offs -- until eventually a formulation of the
equations of motion of the SM without cut-offs is found.

\section{Equilibrium states and the vacuum}
\label{sec:dirac-sea}

How can we tackle problem (II), that is, find manageable approximations of the
in general very complicated dynamics of the Dirac sea model? Clearly, this will
be possible only in special situations, that is, for a certain class of initial
quantum states. Furthermore, such an approximation cannot be carried out without
coarse graining the level of detail of the information that is to be inferred
about the system. 

Consider, as an analogy, a classical gas of $N$
particles confined to the volume $\Gamma$. Although it is in principle possible
to infer the actual motion of the individual particles by solving Newton's law
of motion, for large $N$, this would be a hopeless endeavour and for many
practical purposes unnecessary. For instance, even in equilibrium, the
\emph{microscopic}, actual Newtonian motion of the particles may be very
intricate. However, effectively the net result is that, \emph{macroscopically},
the gas density is almost constant with the variation around this constant
density being small. The smallness requirement depends on the practical purpose
as it determines which effects will be visible and which ones will drown in the
fluctuations. Hence, it defines what is meant by \emph{macroscopic}. In the
same vein, one can introduce further macroscopic parameters like volume,
pressure and temperature. For most engineering purposes their relationship
constitutes a satisfactory description of equilibrated gases, that is, the theory
of thermodynamics. Such a coarse grained description is not restricted to
equilibrium states only. For example, in certain regimes it is possible to
describe the mediation of a perturbation created in the gas by an external
influence in terms of an effective equation for pressure (sound) waves that
excite the initially equilibrated gas, without knowledge of the actual
microscopic Newtonian motion of the individual particles.

Using this classical example as a guide, let us pursue the idea of describing
complex dynamics in terms of small deviations from an equilibrium whose time
evolution can be understood in more simple terms. At first, we only consider
$V(t,x_k)$ in equation \eqref{eq:dirac-eq} to be zero. That is to say, we
stipulate that there are no external influences: the motion of the electrons can
be represented in terms of \eqref{eq:vel-field} and \eqref{eq:dirac-eq}, that is,
subject to the Fermi repulsion (due to the anti-symmetry of $\Psi_t$) and the
Coulomb repulsion only. In analogy to the classical gas, the simple approximate
solutions inferred in the following play the role of the equilibrium states. In
the next section, we then consider the case $V(t,x_k)\neq 0,$ which leads to
excitations of the equilibrium -- the analogue to the pressure waves in the
classical example.

As already indicated, the term creating problem (II) is the Coulomb
pair-interactions encoded in $H^I_k$ given by \eqref{eq:HI}. We therefore define
a class of quantum states for which an approximation is possible in which this
term effectively vanishes and call it the class of equilibrium states. This
class shall consist of all states $\Psi\in\mathcal H^N$ that fulfill the
following two conditions: (a) The quantum mechanical expectation of the
interaction operator $H^I_k$ is approximately constant, that is, there is a
constant $E^I$ such that
\begin{align} 
    \label{eq:HI_const} E^I\approx \langle \Psi, H^I_k(x) \Psi \rangle
    = \Big\langle \Psi, \frac12\sum_{j\neq k} U(x-x_j) \Psi \Big\rangle
    \qquad
    \text{for all }x\in\mathbb R^3\text{ and }k=1,\ldots,N; 
\end{align} 
(b) the fluctuation around this constant expectation value is sufficiently well
behaved. 

In view of the weak law of large numbers, these conditions (a) and (b) can only
be met for sufficiently large $N$. As the theory so far is meant to apply to
the total number of electrons in the universe, $N$ is naturally large. The
exact sense of ``approximatively'' and ``sufficiently'' depends on the
practical purpose: conditions (a) and (b) are there to make sure that the
solutions to the fundamental equation of motion \eqref{eq:dirac-eq} for
$V(t,x_k)= 0$, given an initial state $\Psi_0$ in this class of equilibrium
states, can for all practical purposes be sufficiently well approximated --
e.g. in the sense of reduced density matrices -- by a solution to the much
simpler effective equation of motion 
\begin{align} \label{eq:dirac-approx}
    i\hbar\partial_t \Psi_t^{\approx}(x_1,\ldots,x_N) = \sum_{k=1}^N \left(
    H^0_k(x_k) + E^I \right) \Psi_t^{\approx}(x_1,\ldots,x_N),
\end{align}
replacing the complicated interaction $H^I_k$ by the constant $E^I$.

Regarding the precise mathematical requirements of conditions (a) and (b) and
the precise sense of the approximation we are purposely vague, since the exact
behaviour of the fluctuation needed to carry out the rigorous mathematics is not
entirely settled in the fermionic case (unlike the bosonic case, where for
example Gross-Pitaevski and mean-field approximations can be rigorously
derived). If an initial state $\Psi_0$ fulfills (a) and (b) in a
sufficiently strong sense, the corresponding fully interacting time evolution
\eqref{eq:dirac-eq} is close to the non-interacting one \eqref{eq:dirac-approx}
as the errors which depend on the accumulated fluctuations (b) around the
expectation value (a) can be controlled -- at least in the sense of reduced
density matrices and for large enough $N$ and gas densities. However, while
fermionic $N$-particle states that fulfill (a) are known (e.g., the
non-interacting fermionic ground state, see \eqref{eq:vac-exp} below), it is
unknown whether there are ones that fulfill also condition (b) in a
sufficiently strong sense. In fact, it is conjectured that (b) might be too
strong and that demanding sufficiently small fluctuations for only those
particles with momenta below some threshold might already suffice. This would
mean that condition (b) could even be weakened and we would still be able to
show the closeness of \eqref{eq:dirac-eq} to the approximate time evolution
\eqref{eq:dirac-approx} for such states. The exact notion is however
irrelevant to our discussion. The important point is that
\eqref{eq:dirac-approx} does not have to be assumed, which would be highly
questionable for an interacting Fermi gas, but can be derived from much
simpler and plausible conditions such as (a) and (b).


Before considering a pertinent example of such an equilibrium state, let us put
the motivation for focussing on the class of equilibrium states in other terms.
Assuming that there are only electrons subject to Fermi and Coulomb
repulsion, the ground state \newcommand{\gs}{{\operatorname{gs}}}$\Psi^\gs$ of
such an electron gas is expected to be one in which for almost all initial $Q_0$
with respect to the relevant measure $|\Psi^\gs(X)|^2d^{3N}X$, the electrons are
very homogeneously distributed. Then, if the measure $|\Psi^\gs(Q)|^2d^{3N}X$
gives rise to a homogeneous distribution, our defining condition
\eqref{eq:HI_const} of the class of equilibrium states is a consequence,
and the net effect of $H_k^I$ is a constant potential, canceling out all
interactions: on the level of wave functions, the electrons in such a state effectively do not ``take
notice'' of each others presence. Effectively, they move as if they were
in a vacuum. Hence, the dynamics generated by \eqref{eq:dirac-eq} is very
simple, and, within the bounds of the approximation, it leaves the class of
equilibrium states invariant.

A natural representative for a state in the equilibrium class would be the
actual ground state $\Psi^\gs$ of the interacting system. Its mathematical
structure, however, is extremely complicated and to date not accessible due to
the discussed entanglement induced by the $H_k^I$ terms. Consequently, we have
to find a simpler candidate that replaces $\Psi^\gs$ in the sense of the above
approximation. Physicists usually choose the ground state
$\Psi^\gs_\approx$ of the corresponding approximate equation
\eqref{eq:dirac-approx} given by
\begin{align}
    \label{eq:approx-eigen-eq} \sum_{k=1}^N H^0_k(x_k)\, \Psi_\approx^\gs =
    E_\approx^\gs \, \Psi_\approx^\gs, 
\end{align} 
where $E_\approx^\gs$ is the lowest eigenvalue. Since the
$H^0_k(x_k)$ commute pairwise, $\Psi_\approx^\gs$ can be found by studying the
spectrum of $H^0(x)$ only, which is well-known and given by $\pm
E(p)=\sqrt{p^2+m^2}$ for $p\in \mathcal P_\Lambda=\{k=(k_1,k_2,k_3)\,|\,|k|\leq
\Lambda, k_i=2\pi/Ln_i, n_i\in\mathbb Z,i=1,2,3\}$. Thus, the Hilbert space
$\mathcal H$ splits naturally into the one spanned by the positive energy
eigenstates $\mathcal H^+$ and the negative ones $\mathcal H^-$, respectively.
The ground state $\Psi^\gs_\approx$ is then given by the antisymmetric product state
\begin{align}
    \label{eq:vacuum} 
    \Psi_\approx^\gs = \varphi_{1} \wedge \varphi_{2} \wedge
    \ldots \wedge \varphi_N 
\end{align} 
for $\varphi_n$ denoting any enumeration of the $N$ lowest eigenstates. Among
all such $N$-dependent ground states $\Psi^\gs$, the one for $N = 2|\mathcal
P_\Lambda|$, in which the total number of particles equals twice the number of
admissible momenta (as per $p\in\mathcal P_\Lambda$ there are two spin values),
is distinguished by having the lowest possible value $E^\gs_\approx$. Because
of this fact, one usually considers the case of a sea of $N=2|\mathcal
P_\Lambda|$ many particles, which in physics is referred to as the
\emph{vacuum}. Note that this is merely a convenient mathematical idealization
(and most probably also not exactly fulfilled in nature, as we seem to see more
matter than anti-matter). Furthermore, due to the antisymmetry, one may also choose
$(\varphi_n)_{n=1,\ldots,N}$ to be any other orthonormal basis of the subspace
$\mathcal H^-\subset\mathcal H$, which changes the definition in
\eqref{eq:vacuum} at most by a constant phase factor. As a short-hand notation
we will employ $\Omega=\Psi^\gs_\approx$.

In BM, all there is to the particles comes down to
their position in space only and the evolution of this position as given by a
guiding equation -- in our case equation \eqref{eq:vel-field}. In other words,
in BM, particles are primitive objects in the sense that they only have a
position in space. All the other parameters including mass, charge, spin, etc.\ 
are not additional elements of the ontology characterizing the particles, but dynamical parameters employed
to describe the evolution of the particle positions. Consequently, following BM,
particles are not electrons \emph{per se}, but some particles
are classified as electrons because they move in a certain manner, namely
electronwise so to speak (see \cite{Dickson:2000aa} and \cite{Esfeld:2014}).
Hence, the appearance of negative energy values in the formalism poses no
problem for the Bohmian ontology: an energy value -- be it positive or
negative -- is nothing but a dynamical parameter capturing a particular form of
particle motion. The same goes for positrons and anti-matter in general: all
these are primitive particles moving in a certain manner. Consequently, the fact that in the
vacuum state all the negative energy states are occupied on the level of the
wave functions does not imply that the \emph{physical space} is entirely filled
with particles. Even in the vacuum state there are finitely many particles that
have non-zero distances between them.

It can now be checked directly that for $\Psi=\Omega$ in the equilibrium
condition \eqref{eq:HI_const}, one gets 
\begin{align} \label{eq:vac-exp}
    \eqref{eq:HI_const} = \langle \Omega, H^I_k(x) \Omega \rangle =
    \frac12(N-1)\frac{1}{N} \sum_{n=1}^N \int U(x-y) |\varphi_n|^2(y) \, d^3y,
\end{align}
which is constant (note that $|\varphi_j|=1/|\Gamma|^{1/2}$). This agrees
with Dirac's heuristic picture that, in an equilibrium state, the electrons are
uniformly distributed as, according to Born's rule, the right-hand side of
\eqref{eq:vac-exp} can be interpreted as expectation value of
$\frac12\sum_{j\neq k}U(x-x_j)$ for $(N-1)$ uniformly distributed random
variables $x_j$. Beside this, there are unfortunately only few rigorous results
on the quality of the approximation of \eqref{eq:dirac-approx} to
\eqref{eq:dirac-eq}, that is, whether $\Omega$ fulfills also condition (b) above
to qualify as a state in equilibrium. The indication that it holds in
certain physically interesting regimes comes mainly from the overwhelming
accuracy achieved by predictions obtained using \eqref{eq:dirac-approx} whose
justification requires conditions (a) and (b) as \emph{a priori} assumptions.

In sum, we have defined a class of initial quantum states, namely the
equilibrium class, that allows to solve the in general very complicated dynamics
\eqref{eq:dirac-eq} approximately by the much simpler dynamics
\eqref{eq:dirac-approx}. Furthermore, we have found a particularly simple representative $\Omega$
of this class. Solving the approximate equation
\eqref{eq:dirac-approx} for the initial value $\Psi_0^\approx=\Omega$ yields the
explicit and particular simple evolution $\Psi^\approx_t=\Omega_t$ for 
\begin{align}
    \label{eq:evolved-vacuum} \Omega_t=e^{-i(E_\approx^\gs+N E^I) t}\Omega,
\end{align} 
which can now be taken as approximate solution $\Psi_t$ to equation
\eqref{eq:dirac-eq} for the same initial value $\Psi_0=\Omega$. Such a state
only gives rise to a trivial dynamics of a sea of $N$ electrons in which none of
them ``takes notice'' of the rest as if they were in a ``vacuum'' -- hence the
name \emph{vacuum state}. Nevertheless, this state $\Omega$ will provide the basis for studying more
interesting dynamics in the next section.

Finally, it has to be emphasized that even with $\Psi_t$ and $\Omega_t$ being
close for $\Psi_0=\Omega$ in the sense of reduced density matrices, this
provides little information about the closeness of the velocity fields
\eqref{eq:vel-field} generated by $\Psi_t$ and $\Psi^\approx_t$, respectively.
In general, the respective velocity fields and the corresponding trajectories
differ. However, the results of their statistical analysis agree due to the
approximate agreement of the reduced density matrices. In other words, the price
that one has to pay to overcome the complexity problem (II) above by means of an
approximation in terms of solvable equations is that one has to abandon the hope
of obtaining a calculation of actual trajectories in favour of a statistical
analysis. This is, however, the same situation as in the example of
the classical gas discussed at the beginning of this section. Furthermore, it is
the same situation as in BM: there also is no point in calculating
individual trajectories, since tiny deviations in the initial configuration will
lead to large deviations of the resulting Bohmian trajectories. Consequently,
our knowledge of subsystems of the universe is limited to what can be obtained
from Born's rule, as proven by \citet[][chapter 2]{Durr:2013aa}.

\section{Excitations of the vacuum and the Fock space formalism}
\label{sec:excitations}

More interesting dynamics will take place if we allow $V(t,x_k)$ to be
non-zero, thus including external influences. Under the action of $V(t,x_k)$,
an initial vacuum state $\Omega$, as defined in \eqref{eq:vacuum}, may evolve
into an excited state, as for instance
\begin{align}
    \label{eq:excitation}
    \Phi = \chi \wedge \varphi_2 \wedge \ldots \wedge \varphi_N,
\end{align}
where the element $\varphi_1$ of the orthonormal basis of
$\mathcal H^-$ is replaced by a wave function $\chi\in\mathcal H^+$. For the sake of
simplicity, let us assume that after
the transition from $\Omega$ to $\Phi$, the external influence vanishes again. As
in the preceding section we strive to find an economic effective description of the
time evolution of $\Phi$ to a $\Phi_t$, $t>0$\, that fulfills the complicated dynamics
\eqref{eq:dirac-eq}, that is,
\begin{align}
    \label{eq:}
    i\partial_t \Phi_t = H^N\Phi_t,
\end{align}
where $H^N$ is the $N$-particle Hamiltonian defined in \eqref{eq:dirac-eq}.

In analogy to the pressure waves in the classical gas example mentioned in the
previous section, the leading idea is to describe the complex evolution of
$\Phi_t$ by the evolution of the excitation only as compared to the simple
evolution of the reference state $\Omega_t$ given in
\eqref{eq:evolved-vacuum}. In mathematical terms this can be done by the
following ansatz
\begin{align}
    \Phi_t^\approx(x_1,\ldots, x_N) 
    = 
    \sum_{k=1}^N (-1)^{k+1}\int d^3y_k\, \eta_t(x_k,y_k)
    \Omega_t(x_1,\ldots, y_k, \ldots, x_N).
    \label{eq:ansatz}
\end{align}
The summation and the factor $(-1)^{k+1}$ are to ensure the antisymmetry of the
wave function. For the choice
\begin{align}
    \label{eq:initial-values}
    \Omega_{t=0} = \Omega,
    \qquad
    \eta_{t=0} = \chi \otimes \varphi_1^*,
\end{align}
one obtains $\Phi^\approx_{t=0}=\Phi$ due to the orthonormality of the states
$\chi,\varphi_1,\ldots,\varphi_N$. The excitation is hence encoded by a
two-particle wave function $\eta_t(x,y)$. Its $x$ tensor component, which
we shall refer to as \emph{electron component}, tracks the evolution of the
initial excitation $\chi$. Its $y$ tensor component, which we shall refer to as
\emph{hole component}, tracks the evolution of the corresponding state in the
reference $\Omega_t$ that is missing.

In order to ensure that $\Phi_t^\approx$ approximates the actual $\Phi_t$, it
turns out that one has to choose
\begin{align}
    \label{eq:effective-eta}
    i\partial_t \eta_t(x,y) 
    = 
    H^0_k(x_k)\eta_t(x_k,y_k)
    -\eta_t(x_k,y_k)\overset{\leftharpoonup}{H^0_k}(y_k)
    -U(x_k-y_k)\eta_t(x_k,y_k),
\end{align}
where we set
$\overset{\leftharpoonup}{H^0}(y)=i\alpha\cdot\overset{\leftharpoonup}{\nabla}_y
+ \beta m$ with $\overset{\leftharpoonup}{\nabla}$ denoting the gradient acting
to the left. Hence, the excitation $\eta_t$ evolves according to a two-particle
wave equation obeying the free Dirac dispersion in both tensor components and an
attractive pair-potential potential -- note the signs: the hole component
has positive kinetic energy and the $U$ has a different relative sign as
compared to \eqref{eq:dirac-eq} --, as if it would describe a quantum
two-particles system of identical particles except for the opposite charge.  In
consequence, against the uniform background of the vacuum state $\Omega$,
$\Phi_t$ appears as having one negative elementary charge $e$ and one positive
one.  According to Born's rule, which applies thanks to the discussed mode of
approximation, the position distribution of these two charges is given by
$|\eta_t(x,y)|^2 d^3x d^3y$. In principle, due to the appearance of these two
elementary charges in the transition from $\Omega$ to $\Phi_t$, also the
evolution of all the other tensor components in the sea is affected: their
motion does not exactly follow the one of the true vacuum state $\Omega$ -- a
phenomenon referred to as \emph{vacuum polarization}. This in turn also
influences the evolution of $\eta_t$. However, this effect is not of order $e^2$
as in $U$, but of higher order $e^4$, and thus much smaller. As long as
the excitation comprises only few charges (as compared to $N$), such
polarization effects may be neglected in certain regimes.

In sum, we arrive at an economic two-particle description given by
\eqref{eq:ansatz}, \eqref{eq:initial-values}, \eqref{eq:evolved-vacuum} and
\eqref{eq:effective-eta} that approximates the actual $N$-particle dynamics of
the initial excitation $\Phi$ in \eqref{eq:excitation}. As the initial vacuum
state essentially does not change over time -- cf.\ \eqref{eq:evolved-vacuum} --
and since due to \eqref{eq:HI_const} the vacuum behaves as if there are no
electrons, for all practical purposes, one may be inclined to forget about
$\Omega$ entirely. The vacuum state $\Omega$ then simply encodes a state without
excitations. A transition from $\Omega$ to $\Phi_t$ by means of
\eqref{eq:dirac-eq}, however, requires the introduction of the two-particle wave
function $\eta_t$. Further excitations require the introduction of further
multi-particle excitations and so on. However, as long as the number of
excitations is small compared to $N$, those excitation wave functions are much
less complex objects to keep track off than the actual $N$-particle dynamics
of $\Psi_t$.

To streamline the mathematics according to this idea, it makes sense to
formulate the $N$-particle wave function dynamics \eqref{eq:dirac-eq} not on
$\mathcal H^{\wedge N}$ but on a space that
only keeps track of the varying number of wave function excitations with respect
to a reference state such as $\Omega$. For this purpose one introduces the
so-called Fock space $\mathcal F_\Omega$, which can be defined by employing the
so-called creation and annihilation operators. Algebraically the creation
operator $\psi^*$ is given by
\begin{align}
    \psi^*(f) \; \varphi_1 \wedge \varphi_2 \wedge
    \ldots \varphi_N
    =
    f \wedge \varphi_1 \wedge \varphi_2 \wedge \ldots \varphi_N
\end{align}
for any choice of $f,\varphi_k\in\mathcal H$. The annihilation operator $\psi$
is defined as the corresponding adjoint $\psi$. Due to the anti-symmetric tensor
product $\wedge$ these operators fulfill the anti-commutation relations
\begin{gather}
    \{\psi(f),\psi^*(g)\} = \langle f,g\rangle,
    \qquad
    \{\psi(f),\psi(g)\} = 0
    =
    \{\psi^*(f),\psi^*(g)\}
    \qquad
    \text{ for all } f,g\in\mathcal H,
    \nonumber
    \\
    \text{and furthermore,}
    \quad
    \psi(\chi)\Omega = 0 = \psi^*(\varphi) \Omega
    \qquad
    \text{ for all }
    \chi\in\mathcal H^+, \varphi\in\mathcal H^-.
    \label{eq:relations}
\end{gather}
The Fock space $\mathcal F_\Omega$ can then be defined as the tensor product $\mathcal
F_\Omega = \mathcal F_\Omega^e\otimes \mathcal F_\Omega^h$ of the electron Fock
space $\mathcal F^e_\Omega$ and hole Fock space $\mathcal F^h_\Omega$, which are
spanned by finite linear combinations of the form
$\psi^*(f_1)\ldots\psi^*(f_n)\Omega$ and $\psi(g_1)\ldots\psi(g_n)\Omega$ for
$f_k\in\mathcal H^+$ and $g_k\in\mathcal H^-$, respectively. In this algebraic
construction the specific reference state $\Omega$ in
$\mathcal H^{\wedge N}$ is hidden as $\Omega$ is encoded by  $|\Omega\rangle = 1\otimes
1$. The independence of
$|\Omega\rangle$ of $\Omega$ is, however, an illusion, as $\Omega$ is encoded in
the relations \eqref{eq:relations}  -- acknowledging
the fact that $\Omega$ is the Dirac sea with all states in $\mathcal H^-$
occupied. By construction, there is a well-known canonical isomorphism between $\mathcal
H^{\wedge N}$ and the $N$-particle sector of $\mathcal F^N_{\Omega}$, that is, the
subspace of $\mathcal F_\Omega$ spanned by states with equal numbers of electron
and hole excitations. It reads
\begin{align}
    \iota:\mathcal H^N \to \mathcal F^N_{\Omega},
    \quad
    \iota(f_1 \wedge f_2 \wedge \ldots \wedge f_N) :=
    \psi^*(f_1)\ldots \psi^*(f_N) \psi(\varphi_1) \ldots \psi(\varphi_N)
    |\Omega\rangle
    \label{eq:iso}
\end{align}
for $f_k\in\mathcal H$. Thus,
the approximate excitation $\Phi^\approx_t$ for \eqref{eq:ansatz} is represented in
$\mathcal F_\Omega$ by
\begin{align} 
    \qquad
    |\Phi_t\rangle = \int d^3x\int dy \, \eta_t(x,y) \psi^*(x)\psi(y) |\Omega\rangle.
\end{align}
Using these translation rules implied by $\iota$, we can recast the actual
$N$-particle dynamics \eqref{eq:dirac-eq} generated by the Hamiltonian $H^N$ in
terms of creation and annihilation operators as
\begin{gather}
    i\partial_t |\Psi_t\rangle = \widetilde H |\Psi_t\rangle
    \nonumber
    \\
    \widetilde H
    = 
    \int d^3x\, \psi^*(x) \left( H^0_x + V_t(x) \right) \psi(x) - \int d^3x\int
    d^3y
    \,
    \psi^*(x)\psi^*(y) U(x-y)
    \psi(y)\psi(x).
    \label{eq:dirac-second-quant}
\end{gather}
This expression is the standard second-quantized Hamiltonian of quantum electrodynamics (QED) in Coulomb
gauge when neglecting radiation. Note that in $\widetilde H$ the creation and
annihilation operators appear in pairs, which ensures that for an initial state
in the $N$-particle sector of $\mathcal F_\Omega$ the generated dynamics
also remains there.

In conclusion, for an initial wave function $\Psi_{t=0}\in \mathcal H^{\wedge N}$, there
is a unique initial Fock vector $|\Psi_{t=0}\rangle:=\iota(\Psi_{t=0})$ such
that the time evolutions $(\Psi_t)_{t\in\mathbb R}$ on Hilbert space $\mathcal
H^{\wedge N}$ and $(|\Psi_t\rangle)_{t\in\mathbb R}$ on Fock space $\mathcal
F_{\Omega}$ generated by the Hamiltonians $H^N$ in \eqref{eq:dirac-eq} and
$\widetilde H$, respectively, fulfill $|\Psi_{t}\rangle:=\iota(\Psi_{t})$ for all times $t$. In other words, the $N$-particle formalism is equivalent to
the Fock space formalism when restricting to the $N$-particle sector. While the
former represents the dynamics absolutely, the latter represents the dynamics
with respect to a reference state $\Omega$. The latter is, as we described above for the case of
the two-particle excitation, tailor-made for finding
approximations of excitations of $\Omega$.

\section{The appearance of particle creations and annihilations}
\label{sec:formalism}

In using the second quantized formalism in section~\ref{sec:excitations} instead
of the $N$-particle formalism in \eqref{eq:dirac-eq}, one is naturally led to a
mode of language in which excitations are ``created'' and ``annihilated''. When
an initial reference vacuum state $\Omega$ evolves into an excited state like $\Phi$ in
\eqref{eq:excitation}, an excitation $\eta_t$ is created, which effectively
evolves according to \eqref{eq:effective-eta}. As can be seen from
\eqref{eq:effective-eta}, the interaction between the electron and hole
components of $\eta_t$ is attractive, which may lead to a recombination unless
an external influence $V(t,x)$ prevents it. In such a process the evolved
excitation $\Phi_t^\approx$ decays back to the reference state $\Omega$ resulting in the
annihilation of the excitation.

One may be inclined to refer to the electron component of $\eta_t$ as
describing one created ``electron'' and the hole component as describing one
created ``hole'', or synonymously, ``positron''. This, however, is an abuse of the
well-defined term ``electron'' given by the introduced ontology, namely
a persistent particle that moves electronwise. Even if the
state $\Phi^\approx_t$ in \eqref{eq:ansatz} were not only an approximation but
the actual $N$-particle wave function, by the form of the velocity law
\eqref{eq:vel-field} and antisymmetry, the electron component of $\eta_t$ would
not influence only one electron, but all of them. It is therefore the collective
motion of all electrons that causes the excitation $\eta_t$ to appear. It
would thus be a misconception to interpret $\eta_t$ as guiding a two-particle system.

On the mathematical level of wave functions, the hole component of $\eta_t$
simply encodes what has to be taken out of the reference state $\Omega$, while
the electron component of $\eta_t$ encodes what has to be added to $\Omega$ in
order to get a good approximation of the actual quantum state $\Psi_t$ of the
Dirac sea. This convenient approximation in terms of $\Omega$ comes at the price of losing direct information about the actual particle
trajectories: as mentioned, in this case only the statistics about particle
positions encoded in the wave function are under control. Hence, all that can be
said about $\eta_t$ is that against the uniform distribution of a sea of
electrons described by $\Omega$ -- due to the equilibrium condition
\eqref{eq:HI_const} --, $|\eta_t(x,y)|^2d^3xd^3y$ is the probability of finding
an additional negative charge and the absence of a negative charge (or
mathematically equivalently, the presence of a positive charge) in the volumes
$d^3x$ at $x$ and $d^3y$ at $y$, respectively.  

Nevertheless, the excitation $\eta_t$ has the same properties as wave-packets
and would in principle be capable of guiding two particles. To
emphasize the difference between such excitations and actual particles, the
term \emph{quasi-particles} is usually employed in physics (as, e.g., phonons being the quasi-particles describing quantum excitations of crystal lattice
oscillations). In this view, electron and hole components of $\eta_t$ describe
electron and hole quasi-particles having exactly the same properties except for
the opposite charge, as can be seen from \eqref{eq:effective-eta}. The bottom
line is that at all times there are $N$ electrons in the Dirac sea. None of
them was ever created or destroyed. It is their collective motion, which may
deviate from the vacuum motion, that gives rise to what we call excitations or
electron and hole quasi-particles. 

One can emphasize the quasi-particles character of the excitations already in
the mathematical formalism. For this it is convenient to split $\psi^*,\psi$
into electron and positron creation and annihilation operators 
\begin{align}
    \psi^*(f) = b^*(f) + c^*(f)
    \quad
    \text{with}
    \quad
    b^*(f) = \psi^*(P^+f),
    \quad
    c^*(f) = \psi^*(P^-f)
    f\in\mathcal H,
\end{align}
where $P^\pm$ are the orthogonal projectors on $\cal H$ to $\cal H^\pm$,
respectively.
The number of electrons and hole quasi-particles can then be given by
\begin{align}
    \label{eq:number-operators}
    N^e = \sum_{n} b^*(\chi_n)b(\chi_n),
    \qquad
    N^h = \sum_{n} c(\varphi_n)c^*(\varphi_n),
\end{align}
using any orthonormal bases $(\chi_n)_{n=1,\ldots,N}$ in $\mathcal H^+$ and
$(\varphi_n)_{n=1,\ldots,N}$ in $\mathcal H^-$. Note that by virtue of
\eqref{eq:relations}, for instance,
\begin{align}
    \label{eq:number-op-vac}
   N^e|\Omega\rangle=0=N^h|\Omega\rangle,
\end{align}
while for $\Phi$ in \eqref{eq:excitation}, 
$N^e|\Phi\rangle=1=N^h|\Phi\rangle$ as expected.

However, much caution has to be taken when interpreting the particle number
operators in \eqref{eq:number-operators}. First these operators do not yield a
definite number before taking, for instance, their expectation value. Even then, rather
than an absolute number, these operators record only the number of excitations
relative to the chosen reference state $\Omega$, as can be seen from
\eqref{eq:number-op-vac}, i.e., \eqref{eq:relations}. Furthermore, recall that
in the preceding section when considering the case of zero external influence --
$V(t,x)=0$ --, we argued that the ground state of the free theory, $\Omega$, is
a good candidate to approximate the vacuum. This choice stemmed from the need to
find a good approximation and was by no means unique (which is why we regarded a
whole class of reference equilibrium states -- recall that $\Omega$ was just a
representative in the case of no external influence $V(t,x)=0$). Depending on
the kind of approximation in mind, one choice of reference state $\Omega$ might
be better than another. Changing the reference state, however, naturally changes
the meaning of the number operators, as excitations are defined with respect to
it. 

For instance, when an external influence $V(t,x)$ is present, the
equilibrium conditions -- i.e. fulfilling the conditions (a) and (b) discussed
in \eqref{eq:HI_const} and below -- naturally change. When $V(t,x)$ only changes
very slowly, physicists usually employ the so-called Furry picture: for each time
$t$ one chooses as reference $\Omega$ the state in which all negative energy
states of the Hamiltonian $H^0(x)+V(t,x)$ are filled. However, it is shown
in \cite{fierz_particle_1979} that even this seemingly canonic construction of
reference states \eqref{eq:number-operators} is not Lorentz invariant -- 
in the sense that what might appear as an empty vacuum in one reference frame may
contain many quasi-particles in another. This is because the spectrum, being an
energy, transforms like time under Lorentz boosts. A good choice of a reference
state will ultimately depend on the detector model, namely on which excitations
of the equilibrium conditions \eqref{eq:HI_const} cause clicks. But this is an
artifact of our approximation only, that is, of the attempt in solving the
complexity problem (II) of section \ref{sec:model}.
The Dirac sea model is independent of observers or detector models: there are
always $N$ electrons in the sea moving according to laws
\eqref{eq:vel-field} and \eqref{eq:dirac-eq}.

Consequently, the choice of $\Omega$ should rather be seen as a choice of
suitable coordinates with respect to which one may track the complicated
$N$-particles dynamics \eqref{eq:dirac-eq} most conveniently.  In principle, any
choice is possible, since the isomorphism \eqref{eq:iso} between the
$N$-particle formalism \eqref{eq:dirac-eq} and the formalism in terms of
creation and annihilation operators \eqref{eq:dirac-second-quant} does not
depend on particular properties of $\Omega$. The number operators that are
defined with respect to this choice $\Omega$, however, can be assigned a physical
interpretation only with respect to $\Omega$. For example, if $\Omega$ fulfills the
equilibrium conditions (a) and (b) discussed in \eqref{eq:HI_const} and below,
the net sum of all pair potentials between the electrons vanishes and each
electron effectively feels a constant potential and, thus, moves freely as if it
were in a vacuum (a prime example of dark matter as effectively there is no
interaction except for maybe gravitation). The presence of an excitation of
$\Omega$ disturbs those conditions, resulting in an effective interaction
between all electrons in the sea with the electron-hole excitation
$\eta_t$, which dictates the form of \eqref{eq:effective-eta}. The density
$|\eta_t|^2$ then describes the distribution of an additional negative and
positive charge with respect to the uniform background distribution of $\Omega$.
Only in this sense do the particle number operators record the number of in
principle physically measurable excitations. This is what is gained from the
introduction of an ontology of persisting particles: it is capable of
explaining the import of all the relevant mathematical quantities of the theory.

The Unruh effect (\cite{Unruh:1976}) can also be understood in this way: the
equilibrium conditions were chosen to find a good approximation of the complex
$N$-particle dynamics and are at best Lorentz invariant. Switching to an
accelerated frame will in general violate those conditions. Consequently, states
$\Omega$ and $\Omega'$ describing a vacuum in a rest frame and an accelerated
frame, respectively, may differ. Thus, $\Omega$ may appear in the accelerated
frame as excitation of $\Omega'$ possibly involving many electron-hole
quasi-particle pairs. In sum, there is no point in giving an ontological
significance to these quasi-particles -- which are the ontological
particles in the proposal of
\cite{Bell:1986aa} --,  since there is no canonical choice for $\Omega$.

\section{Conclusion: The merits of the Bohmian approach}

Before drawing our conclusion, three points are worth being addressed. In the first place, 
one may gain the impression that the introduced formalism based on a
reference state $\Omega$, which fills the spectral subspace $\mathcal H^-$ with
actual particles that move electronwise and leaves $\mathcal H^+$ empty, violates
one of the much celebrated symmetries of QED, namely the so-called charge
symmetry (a symmetry that is broken in the weak interaction). This is a
misconception. Charge symmetry simply demands that the equations of motion
remain invariant when the signs of the elementary charge $e$ and the one of the
external influence are inverted, which is the case (see \eqref{eq:dirac-eq} and
also \eqref{eq:effective-eta} for the effective equation).

Furthermore, the model that we have discussed is based on a finite, though large, number
of electrons. Whenever one or both of the parameters $\Gamma$ and $\Lambda$ --
the volume of the universe and the cut-off of the interaction -- are sent to
infinity, the construction of the ground state in \eqref{eq:vacuum} would imply
$N\to\infty$, as then $\mathcal H^-$ comprises countable but infinitely many
states. Sending $\Gamma$ to infinity could on the one hand be seen as carrying
out a thermodynamic limit of the finite system. On the other hand, even keeping
the volume $\Gamma$ fixed while sending the momentum cut-off $\Gamma$ to
infinity would as well require $N\to\infty$. In the latter case, the only way to
maintain a sensible ontology of particles is to assume that the Dirac sea is not
entirely filled. For instance, \eqref{eq:vacuum} may be composed only of the
first $N$ states in $\mathcal H^-$ occupied by electrons. When coupling such a
state to an open system such as the photon field, the electron in the Dirac sea
may decay to lower and lower energy states that are unoccupied. This would
result in a never ending energy transfer from the electron into the photon
degrees of freedom that transport this energy to spatial infinity so that the
dynamics becomes unstable. Such a situation would of course be unphysical.
However, it is due to the unphysical artifact of the photon field acting as an
infinite energy sink. In a realistic model of the universe such an energy sink
has to be ruled out. To achieve this, the equations of motion of the photon
field would have to be supplied with an extra condition ensuring that all
emitted photons will at some later time be absorbed again by fermions (see
\cite{Wheeler-Feynman:1945aa}) so that no radiation can escape to spatial
infinity. For the moment, however, an ontological discussion taking into
account the removal of the cut-offs as well has to wait until modern QFT
succeeds in finding a well-defined, relativistically interacting equation of
motion that will replace equation \eqref{eq:dirac-eq}.

Finally, it has to emphasized that even in the non-interacting case, while the
statistics about the particle positions (which are encoded in the wave functions
of the excitations thanks to Born's law) are Lorentz invariant, the actual
particle trajectories generated by the velocity law \eqref{eq:vel-field} are
not. It is a well-known fact that the latter depend on a preferred foliation of
Minkowski space-time -- in the setup above, we have chosen equal time
hypersurfaces. This circumstance is not an artifact of our model; it is
generic to all relativistic versions of BM. The reason is the
manifestly non-local nature of QM as demonstrated, for instance, by Bell's theorem and the subsequent experiments. Instead of
adding this extra structure of a preferred foliation by hand, it is possible to introduce it in
a Lorentz-covariant manner by taking it to be determined by the $N$-particle wave function
itself, or equivalently the Fock space vector (see \cite{Duerr:2013}). We did not do so because this would have made the
definition of the Dirac sea model and our arguments for a particle ontology
unnecessarily opaque. Furthermore, the actual trajectories of individual electrons figure only in the ontology of our theory, while, due to problem (I) mentioned in section \ref{sec:model}, they alone do
not allow to infer predictions. It is rather the collective motion of all
particles in the Dirac sea that generates the correct position statistics, which
are encoded in the excitation wave functions by Born's law; the latter are
accessible and Lorentz-invariant.

Despite the still open desiderata, we have demonstrated that the predictions
of the SM in the electron sector with cut-offs and neglect of radiation arise
naturally from a theory of $N$ persistent particles.  Creation and annihilation
appear only in an effective description with respect to a chosen reference
state. This programme is not only applicable to the electron, but to all fermion
sectors.

In conclusion, let us stress again the parallelism with BM for QM: in both
cases, the ontology is exactly the same -- a fixed, finite number of permanent
point particles that are characterized only by their positions -- , and the
dynamics is of the same type, that is, a deterministic law describing the motion
of the particles on continuous trajectories without any jumps. In both cases,
the objective is to answer the questions of what there is in nature on the
fundamental level (i.e. permanent particles) and how what there is evolves (i.e.
provide a dynamics for the individual processes in nature). From this ontology
and dynamics then follow the formalisms of statistical predictions of both
standard QM and QFT. In this sense, these predictions are explained by this
underlying ontology and dynamics.

The same goes for the solution to the measurement problem that hits QFT in the
same way as QM (see \cite{Barrett:2014aa}): the Bohmian point particles are
admitted because they explain the measurement outcomes. For instance, the dots on
the screen as recorded in the double slit experiment are made up by these point
particles. However, this explanation is not achieved by the (primitive)
ontology of particles alone, but by this ontology together with the dynamics
(i.e. the guiding equation): it is the dynamics that provides for the stability
of macroscopic objects, including pointer positions and dots on a screen. As
\cite{Dickson:2000aa} convincingly argued, to obtain this stability, no
properties of the particles over and above their position in space are needed,
but a dynamics is that provides for trajectories such that there are stable
macroscopic particle configurations.

Bearing these facts in mind, the solution to the measurement problem
provided by Dirac sea Bohmian QFT is of the same type: here again, the Bohmian
point particles are admitted because they explain the measurement outcomes.
There always is a definite, finite number of particles moving
on continuous trajectories, with some of these particles making up the
macroscopic phenomena that we see. As discussed, we only have to be more
careful to relate those particles and their dynamics to what we see with the
naked eye, e.g., the trajectories recorded in cloud chamber detectors. The
latter are generated by deviations from the collective motion of the Bohmian
particles in an equilibrium state such as the vacuum.  Thus, to stress again,
what explains these phenomena is not the mere fact of there being particle
configurations, but the particle dynamics: it consists in this case of 
excitations from a sea of particles in an equilibrium state, with these
excitations, as elaborated on above, implying a change of the motion of in
principle all the particles in the sea.  To sum up, what explains the traces in
the cloud chamber are these excitations as affecting the motion of the
particles in the sea.  

Hence, whatever may eventually be the fate of the Bohmian approach, it
works for QFT in the same way as for QM. Thus, if this approach is a serious
contender for the ontology of QM, so it is a serious contender for the ontology
of QFT.

\paragraph{Acknowledgement.} We are grateful to J. Barrett, D. Dürr, S.
Goldstein, T. Maudlin, W. Struyve, N. Zanghì and all the participants of the
summer school ``The Ontology of Physics'' in Saig (Black Forest, Germany) in
July 2015 for discussions on the proposal set out in this paper. D.-A. Deckert
thanks M.  Jeblich, D. Mitrouskas, S.  Petrat and P. Pickl for valuable
discussions about the mathematical control of the fluctuations of the particles
in a Fermi gas that is the content of their forthcoming article and also let to
point (b) in the definition of the equilibrium states above. Furthermore, D.-A.
Deckert is grateful to D. Dürr and F.  Merkl for many physical and mathematical
insights and enlightening exchange on the Dirac sea. A. Oldofredi's work on
this paper was supported by the Swiss National Science Foundation, grant no.
105212-149650, while D.-A.\ Deckert's work was funded by the junior research
group grant \emph{Interaction between Light and Matter} of the Elite Network of
Bavaria.

\end{document}